\documentclass[letterpaper]{article} 
\usepackage{aaai2026}  
\usepackage{times}  
\usepackage{helvet}  
\usepackage{courier}  
\usepackage[hyphens]{url}  
\usepackage{graphicx} 
\usepackage{natbib}  
\usepackage{caption} 
\usepackage{algorithm}

\usepackage{amsmath}
\usepackage{algpseudocode}
\usepackage{multirow}
\usepackage{booktabs}
\usepackage{subfigure}
\usepackage{subcaption}
\usepackage{tabularx}
\usepackage{amssymb}

\usepackage{newfloat}
\usepackage{listings}
\DeclareCaptionStyle{ruled}{labelfont=normalfont,labelsep=colon,strut=off} 
\floatstyle{ruled}
\newfloat{listing}{tb}{lst}{}
\floatname{listing}{Listing}
\title{FineRef: Fine-Grained Error Reflection and Correction for \\
Long-Form Generation with Citations}
\author{
    Yixing Peng\textsuperscript{\rm 1,2}, Licheng Zhang\textsuperscript{\rm 1}\thanks{Corresponding author}, Shancheng Fang\textsuperscript{\rm 3}, Yi Liu\textsuperscript{\rm 2}, Peijian Gu\textsuperscript{\rm 1}, Quan Wang\textsuperscript{\rm 4}
}
\affiliations{
    \textsuperscript{\rm 1}University of Science and Technology of China\\
    \textsuperscript{\rm 2}State Key Laboratory of Communication Content 
Cognition, People's Daily Online\\
    \textsuperscript{\rm 3}Shenzhen University \qquad
    \textsuperscript{\rm 4}Beijing University of Posts and Telecommunications

    {\{xk98, zlczlc, gpj123\}}@mail.ustc.edu.cn,  fangsc@szu.edu.cn, gavin1332@gmail.com, wangquan@bupt.edu.cn
}

\usepackage{bibentry}

\begin{document}

\maketitle

\begin{abstract}
Generating with citations is crucial for trustworthy Large Language Models (LLMs), yet even advanced LLMs often produce mismatched or irrelevant citations. Existing methods over-optimize citation fidelity while overlooking relevance to the user query, which degrades answer quality and robustness in real-world settings with noisy or irrelevant retrieved content. Moreover, the prevailing single-pass paradigm struggles to deliver optimal answers in long-form generation that requiring multiple citations. To address these limitations, we propose \textbf{FineRef}, a framework based on \textbf{Fine}-grained error \textbf{Ref}lection, which explicitly teaches the model to self-identify and correct two key citation errors—mismatch and irrelevance—on a per-citation basis. FineRef follows a two-stage training strategy. The first stage instills an “attempt–reflect–correct” behavioral pattern via supervised fine-tuning, using fine-grained and controllable reflection data constructed by specialized lightweight models. An online self-reflective bootstrapping strategy is designed to improve generalization by iteratively enriching training data with verified, self-improving examples. To further enhance the self-reflection and correction capability, the second stage applies process-level reinforcement learning with a multi-dimensional reward scheme that promotes reflection accuracy, answer quality, and correction gain. Experiments on the ALCE benchmark demonstrate that FineRef significantly improves both citation performance and answer accuracy. Our 7B model outperforms GPT-4 by up to 18\% in Citation F1 and 4\% in EM Recall, while also surpassing the state-of-the-art model across key evaluation metrics. FineRef also exhibits strong generalization and robustness in domain transfer settings and noisy retrieval scenarios.
\end{abstract}



\section{Introduction}

Large Language Models (LLMs) \cite{brown2020language, achiam2023gpt} excel at generating long-form response to user queries, yet they are often prone to hallucinations that producing factually incorrect content. Retrieval-Augmented Generation (RAG) \cite{borgeaud2022improving, izacard2023atlas} has emerged as a promising solution by grounding responses in external passages, significantly improving factuality and knowledge coverage. Building on this paradigm, generation with citations has become a core requirement for trustworthy language systems \cite{gao2023enabling, yue2023automatic}. By explicitly requiring the model to attach citations in its responses to the supporting retrieved passages, this paradigm (1) enhances answer verifiability by enabling users to trace the sources underlying the model's output, and (2) constrains the generation to more faithfully align with the retrieved evidence, thereby mitigating hallucinations. However, advanced LLMs and commercial chat engines still often produce unsupported or inaccurate citations in practice \cite{liu2023evaluating, gao2023enabling}, which undermines the credibility of their outputs and limits their applicability in high-trust domains such as healthcare, law, and scientific research.

\begin{figure}[t]
\centering
\includegraphics[width=1\columnwidth]{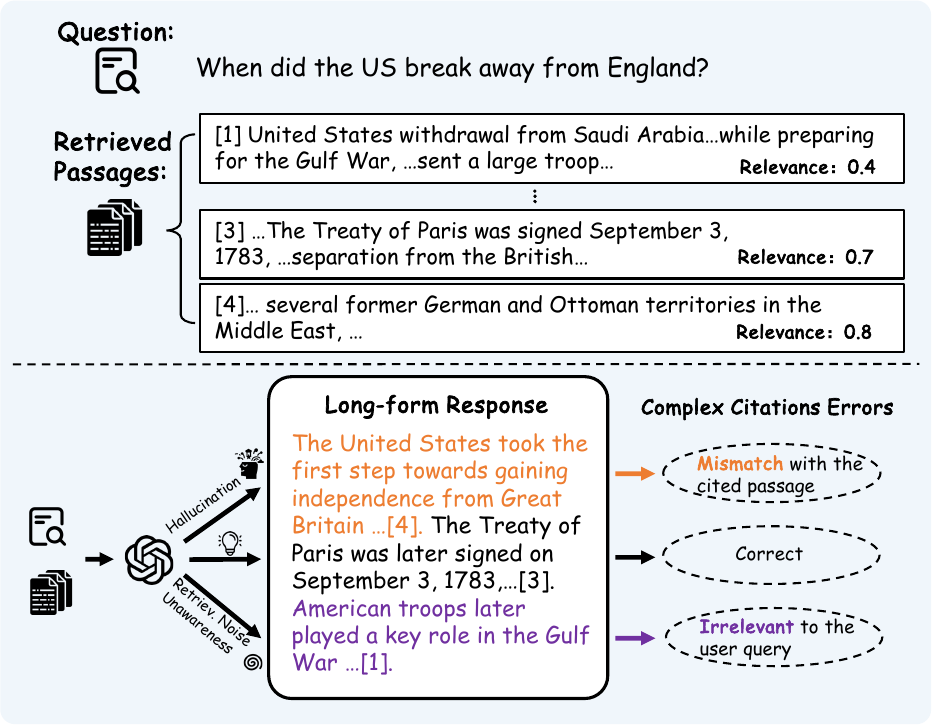} 
\caption{An example of citation errors in generated response: orange indicates citations that do not match the referenced passage (mismatch), while purple denotes citations that are irrelevant to the query (irrelevance).}
\label{intro}
\end{figure}

Progress on these challenges has been hindered by annotation difficulty. While several efforts have explored training-free methods such as in-context learning \cite{gao2023enabling,li2023helma} and Chain-of-Thought (CoT) \cite{ji2024chain}, their performance remains limited. 
Differently, the work represented by \cite{aly2024learning} adopts iterative self-training by selecting citation-accurate samples from the model's own outputs. Nevertheless, these approaches still suffer from several key limitations. 
First, existing work typically trains only on citation-accurate samples, which helps mitigate mismatched citations—generating statements that are not supported by the attributed sources. However, this implicitly assumes that all retrieved passages are relevant to the query, limiting the model’s ability to identify irrelevant content. 
In real-world scenarios, retrieval results often contain noisy or unrelated passages, leading to irrelevant citations, where the model cites passages unrelated to the query. This distribution gap between clean training data and noisy application scenarios significantly weakens model robustness and answer quality.
Second, prevailing approaches adopt a single-pass generation paradigm, where the model is required to produce a fully cited response in one step. 
While this simplifies the generation process, it is inadequate for complex real-world scenarios, particularly long-form generation tasks that demand multiple, contextually appropriate citations and are prone to diverse citation errors.
Without mechanisms for planning or revision, the model struggles to ensure citation correctness within a single pass.

To address these challenges, a promising solution is to introduce a self-reflection mechanism that enables the model to proactively identify and revise citation errors after generation. Prior work in other domains \cite{madaan2023self, shinn2023reflexion} typically adopts coarse-grained, answer-level reflection. However, in long-form generation tasks involving multiple citations, such coarse-grained reflection makes it difficult to locate erroneous citations and distinguish error types. Moreover, when reflection signals are generated directly by general-purpose LLMs, they often exhibit inconsistency in both accuracy and structure, as distinct decisions can be made for the same input, making them unreliable as training supervision. Therefore, there is a pressing need to develop a fine-grained, controllable reflection framework to support precise error identification and correction.

\begin{figure*}[t]
\centering
\includegraphics[width=0.95\textwidth]{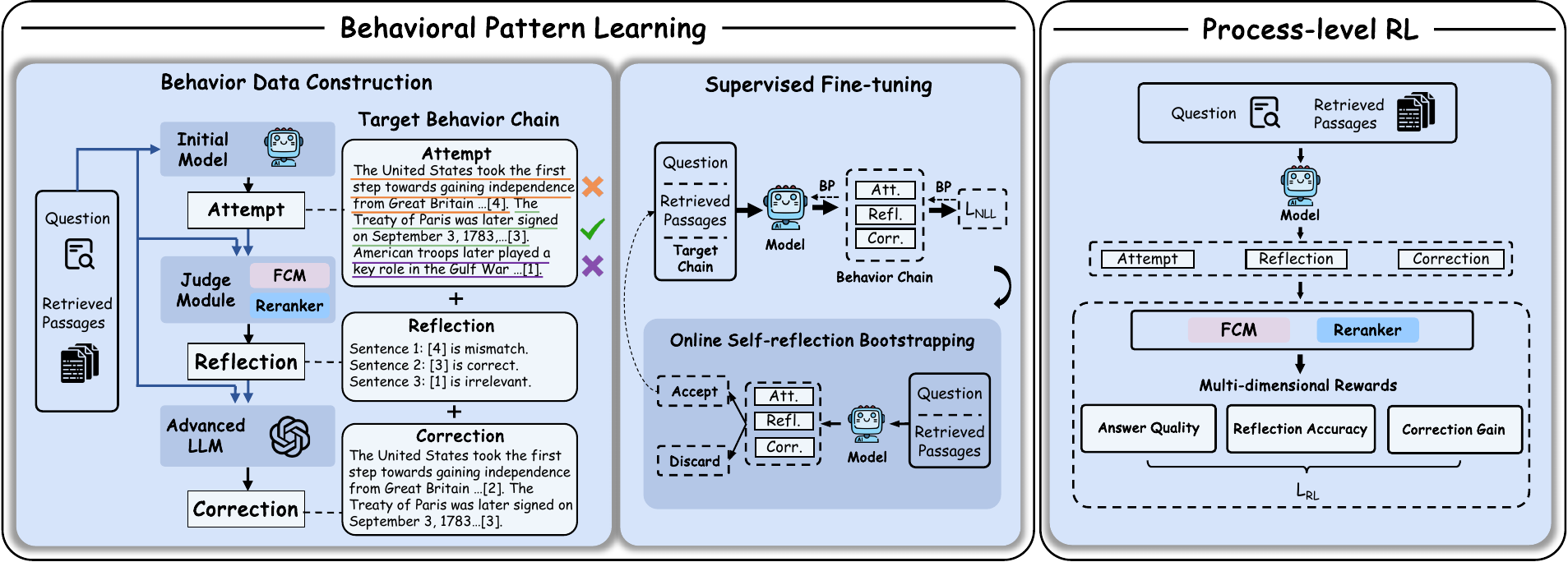} 
\caption{FineRef involves two training stages: (1) \textit{Behavior pattern learning} stage, where the model is supervised to generate the “attempt–reflection–correction” chain using fine-grained reflection data constructed via specialized FCM and reranker models, followed by online reflection bootstrapping, improving from self-generated reflection-correction data. (2) Process-level RL stage, a multi-dimensional reward function further enhances answer quality, reflection accuracy, and correction effectiveness.}
\label{method}
\end{figure*}

To this end, we propose a framework for generation with citations based on {\bf Fine}-grained error {\bf Ref}lection ({\bf FineRef}), which endows the LLM with citation-level self-reflection capabilities, enabling it to explicitly identify and correct two common and impactful citation errors - mismatch and irrelevance - from its own initial attempt. FineRef adopts a two-stage training strategy, as shown in Figure~\ref{method}, including (1) \textbf{behavioral pattern learning} that apply supervised fine-tuning to teach the attempt–reflect–correct behavioral pattern for citation generation, (2) \textbf{process-level reinforcement learning} to enhance the model’s reflection and correction capabilities. In the first stage, to construct behavioral data, we first prompt the initial model attempt to generate response with citations. We then leverage a specialized yet lightweight factual consistency model (FCM) \cite{kryscinski2020evaluating} and a reranker \cite{bge_embedding} to identify mismatch and irrelevance errors for each citation in the model-generated attempts and automatically construct fine-grained, accurate reflection signals. Correction data is generated using advanced LLM based on these signals, and high-quality attempt–reflection–correction chains are subsequently constructed and curated for training. To improve generalization, we propose an \textbf{online self-reflective bootstrapping strategy} where the model generates full attempt–reflect–correct chains during training, and retains those that successfully identify and fix citation errors, exposing the model to diverse reflection patterns and enhancing its adaptability to complex citation scenarios. In the second stage, we conceptualize the generation process as multiple sub-behaviors and design a \textbf{multidimensional reward scheme} given the distinct objectives of each sub-behavior. For reflection behavior, we introduce a reward for reflection accuracy, encouraging the model to correctly identify citation errors. For the attempt and correction behaviors, we apply rewards for citation and answer quality, and further incorporate a correction gain reward to incentivize improvements of the correction over the initial attempt. This approach (1) overcomes the limitations of existing methods that focus solely on citation fidelity, and enhances citation quality while preserving overall question-answering performance. (2) Through this reflect-correct process, our framework significantly improves the robustness in complex citation scenarios.

We conduct experiments on ALCE \cite{gao2023enabling}, a few-shot benchmark for citation generation, which includes two challenging long-form answering datasets: ASQA and ELI5. We evaluate using different LLMs. Based on only four annotated examples, our 7B model achieves up to an 18\% improvement in Citation F1 over GPT-4 \cite{achiam2023gpt}, while also enhancing answer performance by up to 4\% in EM Recall. Moreover, FineRef significantly outperforms the state-of-the-art model CALF \cite{aly2024learning} across a range of backbone LLMs. Furthermore, consistently strong performance of our method under domain transfer and noisy retrieval conditions highlights its generalization and robustness capabilities.

\section{Task Formulation}
Given a question $q$ and a set of retrieved passages $P = \{p_1, \dots, p_m\}$, the goal is to generate a long-form answer $\hat{y} = \{s_1, \dots, s_n\}$. In long-form answer generation with citations, each generated sentence $s_i$ is expected to cite one or more relevant passages $C_i \subseteq P$ (denoted by bracketed indices, e.g., “[1]” referring to $p_1$), such that the content of $s_i$ is entailed by the cited passages $C_i$. The task requires the information in $s_i$ to originate from the attributed/cited passages $C_i$, ensuring that $\hat{y}$ is fully verifiable by $P$. Moreover, the overall answer $\hat{y}$ is required to be factually correct and responsive to the question $q$, going beyond citation fidelity to ensure the answer's informativeness and relevance. 

\section{Method}
FineRef is a framework for generation with citations that models the process of identifying and correcting citation errors through fine-grained self-reflection.  As illustrated in Figure~\ref{method}, FineRef employs a two-stage training pipeline: (1) \textbf{Behavioral pattern learning}, which instills the attempt–reflect–correct behavior via supervised learning, and (2) \textbf{Process-level reinforcement learning}, which optimizes using a multi-dimensional reward to promote reflection accuracy, answer quality, and correction gain.

\subsection{Behavioral Pattern Learning}
\paragraph{Behavior Data Construction.} To train a citation generation model $F$ capable of reflecting on and correcting its own citation errors, we first construct structured training data comprising three components: \textbf{attempt}, \textbf{reflection}, and \textbf{correction}. The \textit{attempt} is designed to simulate the model’s initial citation behavior. Specifically, given a question $q$ and a set of retrieved passages $P = \{p_1, \dots, p_m\}$, we prompt the model $F$ to generate an answer with citations $y_c$, and use it as the initial attempt.

For \textit{reflection} construction, we move beyond relying on coarse-grained reflections directly generated by LLMs, which often lacks precision, fails to localize erroneous citations, and cannot differentiate between error types, making it unsuitable for supervised correction training. Instead, we adopt a fine-grained and controllable reflection construction strategy, grounded in explicit error typing.

To ensure the reliability of reflection labels, we adopt a precision-oriented labeling pipeline. For \textbf{mismatch} errors, we employ a FCM \cite{zha2023alignscore} to evaluate the alignment between the claim and corresponding attributed passages. Specifically, for each sentence $s_i$ and its attributed passage \( c_{ij} \), we computes:
\[
o_{ij} = \phi(s_i, c_{ij}), \quad o_{ij} \in [0, 1]
\]
where $\phi$ denotes the FCM and $o_{ij}$ is the predicted consistency score. A predefined threshold is applied to determine whether a claim (the sentence) is inconsistent with its source (the attributed passage) (i.e., mismatch).

To detect \textbf{irrelevant} citations those unrelated to the input question $q$ despite support the claim, we apply a reranker \( \gamma(s_i, q) \in \{0,1\} \), which makes a binary judgment on whether each citation $c_{ij}\in C_i$ of the sentence $s_i$ is relevant to the question. Each citation is categorized as follows:
\begin{itemize}
    \item \textbf{Mismatch}: if the consistency score is below the threshold;
    \item \textbf{Irrelevance}: if not mismatch and the reranker predicts irrelevance;
    \item \textbf{Correct}: otherwise.
\end{itemize}

These error type labels are aggregated into structured, sentence-level reflection annotations. Compared with coarse-grained LLM-generated feedback, this method provides \textbf{fine-grained and controllable supervision}, enabling more effective learning of reflection behavior.

For the \textit{correction}, we adopt in-context learning approach, using a strong instruction-aligned LLM (e.g., GPT-4o) to revise the initial attempt based on its corresponding reflection. To ensure quality, we apply dual filtering based on two metrics: (1) a citation quality score $Q_{\text{cite}}(correction) = \text{CitationF1}(correction, C, \phi)$, computed based using an FCM, and corresponds to the harmonic mean of citation precision and recall\footnote{Metric details can be referred to in Appendix.}. (2) An answer quality evaluation score, $Q_{\text{ans}}(correction) = \text{Correctness}(correction, A)$ measures the degree of alignment between the generated answer and the reference answer $A$ (e.g., using Exact Match score). We design an acceptance function to determine whether a correction instance should be accepted. 
\begin{equation}
\text{Accept}(\hat{y}^c, \hat{y}^a) = 
\mathbb{I} \left[ 
\begin{aligned}
& Q_{\text{cite}}(\hat{y}^c) \geq \tau_{\text{cite}} \wedge 
Q_{\text{ans}}(\hat{y}^c) \geq \tau_{\text{ans}} \\
& \wedge \Big( Q_{\text{cite}}(\hat{y}^c) > Q_{\text{cite}}(\hat{y}^a) \\
& \wedge Q_{\text{ans}}(\hat{y}^c) > Q_{\text{ans}}(\hat{y}^a) \Big)
\end{aligned}
\right]
\end{equation}
where $\mathbb{I}$ is the indicator function. Only when $\text{Accept}(\hat{y}^c, \hat{y}^a)=1$, correction $\hat{y}^c$ can be retained.

\paragraph{Initial Training.} After data construction, we combine the attempt, reflection and correction into a complete \textit{attempt--reflection--correction} chain $y$ and use them to perform supervised fine-tuning on model $F$. This stage enables the model to learn the behavior of explicitly identifying and correcting citation errors based on structured reflection signals. Joint optimization over the attempt–reflect–correct chain promotes internal consistency between reflection and correction behaviors. Specifically, at the initial stage, we warm up the model by minimizing the negative log-likelihood loss:
\begin{equation}
\mathcal{L}_{NLL} = - \frac{1}{|y|}\sum_{t=1}^{|y|}\text{log} p_\theta(y_t | q, P, y_{<t})
\end{equation}

\paragraph{Online Self-Reflective Bootstrapping.} 
In order to enhance generalization in complex citation scenarios, we propose an \textit{online elf-reflective bootstrapping} strategy. After the initial warm up, the model iteratively generates \textit{attempt--reflection--correction} chains. We retain those chains for which $\text{Accept}(y^{c}, y^{a}) = 1$, and incorporate them into the training set for the subsequent training epochs. To maintain high-quality reflection supervision, we replace the model-generated reflections with more accurate ones, automatically constructed using the same FCM and reranker model from the data construction pipeline. This self-sampling and training process enables the model to continually learn from self-generated samples where citation errors can be accurately reflected and corrected.

\subsection{Process-Level RL with Multi-Dimensional Rewards}
After the first-stage learning, the model has acquired a preliminary ability to generate with citations and perform self-reflective correction. To further strengthen this capability, we adopt reinforcement learning (RL) to optimize the model behavior across the entire \textit{attempt--reflection--correction} process. Inspired by process-level learning \cite{shao2024deepseekmath}, we treat this generation process as a composition of interdependent sub-behaviors and design a multi-dimensional reward function to supervise each sub-behavior.

Our reward design consists of three components. First, we define a \textbf{reflection accuracy reward} to encourage correct classification of citation error types. For $N$ citations, let $\hat{y}_i$ denote the predicted error type for citation $i$ and $y_i$ represents the ground-truth (automatically derived as reflection construction). The reward is computed as:
\begin{equation}
R_{r}^{\text{refl}}(r|x,y_{:r}) = \frac{1}{N} \sum_{i=1}^{N} \left( \mathbb{I}[\hat{y}_i = y_i] - \mathbb{I}[\hat{y}_i \neq y_i] \right)
\end{equation}

Second, we assign a binary \textbf{citation and answer quality reward} to both the attempt and correction behaviors:
\begin{equation}
\begin{aligned}
R_{c}^{\text{ans}}(c|x,y_{:c}) = &
\begin{cases}
+1, & \text{if } Q_{\text{cite}}(y^{corr}) \geq \tau_{cite} \\
& \wedge Q_{\text{ans}}(y^{corr}) \geq \tau_{ans} \\
-1, &otherwise
\end{cases} \\
R_{a}^{\text{ans}}(a|x,y_{:a}) = &
\begin{cases}
+1, & \text{if } Q_{\text{cite}}(y^{attempt}) \geq \tau_{cite}^{attempt} \\
& \wedge Q_{\text{ans}}(y^{attempt}) \geq \tau_{ans}^{attempt} \\
-1, &otherwise
\end{cases}
\end{aligned}
\end{equation}
where $\tau_{cite}^{(\cdot)}$ and $\tau_{ans}^{(\cdot)}$ denote predefined thresholds.

Finally, to encourage meaningful correction beyond the initial attempt, we introduce a \textbf{correction gain reward}, computed based on the improvement in generation quality:
\begin{equation}
\begin{aligned}
R_{c}^{\text{gain}}(c|x,y_{:c} )= &
\begin{cases}
+1, & \text{if } Q_{\text{cite}}(y^{corr}) \geq Q_{\text{cite}}(y^{attempt}) \\
& \wedge Q_{\text{ans}}(y^{corr}) \geq Q_{\text{ans}}(y^{attempt}) \\
-1, &otherwise
\end{cases}
\end{aligned}
\end{equation}
This reward formulation enables fine-grained credit assignment across the entire citation generation process and effectively guides the model toward more accurate and self-correctable behaviors.

Drawing inspiration from the process-level GRPO paradigm \cite{shao2024deepseekmath}, all behaviors of the same type (i.e., attempt, reflection, or correction) belong to a group. The unified reward context of the behavior in a group is defined as follows:
\begin{equation}
\mathbf{R}(z_t) = \left( R_{z_i}(z_i \mid x, y_{:z_i}) \right)_{i=1}^{t-1}
\end{equation}
where $y_{:z_i}$ denotes the sequence of behaviors preceding the behavior $z_i$.
Following REINFORCE with Leave-One-Out (RLOO) \cite{ahmadian2024back}, we compute the advantage for each behavior in a group:
\begin{equation}
\begin{aligned}
A_{z_t}(x,y) = &R_{z_t}(x,y_{:z_t}) - b_{z_t}(x,y) \\
&-\beta log \frac{\pi _{\theta_{old}}(z_t | x,y)}{\pi _{ref}(z_t|x,y)}
\end{aligned}
\end{equation}
where the $b_{z_t}(x,y)$ is baseline, which can be computed as follows:
\begin{equation}
\begin{aligned}
&\quad b_{z_t}(x,y) = \frac{1}{|G(\mathbf{R}(z_t))|} \sum_{z\in G(\mathbf{R}(z_t))} R_z(z_t|x^{(z)},y_{:z_t}^{(z)})
\end{aligned}
\end{equation}
where $G(\cdot)$ denotes the group. This baseline estimation reduces the variance of the policy gradient and improves training stability. The final policy gradient loss is computed as:
\begin{equation} 
\begin{aligned}
\mathcal{L}_{RL} = &- E_{x\sim D,y\sim \pi _{\theta_{old}(\cdot |x)}} [\frac{1}{|y|_z}\sum_{z\in y}min(r_z(\theta) A(z|x,y_{:z}),\\
&clip(r_z(\theta), 1-\epsilon,1+\epsilon) A(z|x,y_{:z}))]
\end{aligned}
\end{equation}
where $r_z(\theta)=\frac{\pi_\theta(z|x,y_{:z})}{\pi _{\theta_{old}}(z|x,y_{:z})}$ is the importance ratio, $\theta$ is the parameters of the model $F$.

\begin{table*}[ht]
\centering
\setlength{\tabcolsep}{1mm}
\begin{tabular}{l|ccccc|ccccc}
\toprule
\multirow{3}{*}{\textbf{Method}} & \multicolumn{5}{c|}{\textbf{ALCE-ASQA}} & \multicolumn{5}{c}{\textbf{ALCE-ELI5}} \\
& Similarity & Fluency & Correct & Correct & Citation & Similarity & Fluency & Correct & Correct & Citation \\
& Rouge-L & MAUVE & EM Rec. & . in P & $F_1$ & Rouge-L & MAUVE & EM Rec. & . in P & $F_1$ \\
\midrule
ChatGPT & -- & 66.6 & 40.4 & -- & 73.1 & -- & 57.2 & 12.0 & -- & 50.5 \\
GPT-4 & -- & 67.1 & 41.3 & -- & 71.9 & -- & 38.4 & 14.2 & -- & 46.9 \\
AGREE & -- & -- & 40.9 & -- & 75.1 & -- & -- & - & -- & - \\
Self-RAG 7B & 35.7 & 74.3 & 30.0 & -- & 67.3 & 16.9 & 32.6 & 9.7 & 5.4 & 27.6 \\
BP, T5-3B & -- & -- & 33.8 & -- & 77.8 & -- & -- & 5.2 & - & 60.9 \\

\midrule
\multicolumn{11}{c}{\textbf{LLaMA2-7B-chat}} \\
\midrule
In-context & $35.9_{0.3}$ & $77.8_{3.1}$ & $35.0_{0.6}$ & $25.7_{0.6}$ & $49.9_{1.0}$ & $20.5_{0.2}$ & $36.2_{2.5}$ & $17.7_{0.6}$ & $10.8_{0.6}$ & $38.2_{0.6}$ \\
Few-shot FT & $34.9_{0.4}$ & $69.2_{4.3}$ & $32.0_{0.4}$ & $22.3_{0.7}$ & $55.0_{1.8}$ & $\underline{21.3}_{0.2}$ & $58.2_{2.2}$ & $\underline{17.8}_{0.6}$ & $11.2_{1.1}$ & $48.7_{2.9}$ \\
CaLF & $\underline{37.8}_{0.4}$ & $\textbf{86.0}_{3.7}$ & $\underline{37.7}_{0.6}$ & $\underline{29.3}_{0.4}$ & $\underline{70.4}_{2.5}$ & $20.8_{1.0}$ & $\underline{59.6}_{11.5}$ & $17.0_{0.3}$ & $\underline{11.9}_{0.2}$ & $\underline{66.5}_{5.9}$ \\
Ours & $\textbf{38.5}_{0.3}$ & $\underline{85.2}_{2.9}$ & $\textbf{40.3}_{0.5}$ & $\textbf{32.1}_{0.8}$ & $\textbf{74.9}_{1.9}$ & $\textbf{21.5}_{0.2}$ & $\textbf{60.1}_{4.4}$ & $\textbf{19.2}_{0.7}$ & $\textbf{13.5}_{0.3}$ & $\textbf{68.4}_{2.7}$ \\
\midrule
\multicolumn{11}{c}{\textbf{Mistral-Orca-7B}} \\
\midrule
In-context & $38.7_{0.1}$ & $54.7_{1.8}$ & $40.2_{0.3}$ & $31.9_{0.2}$ & $55.6_{0.8}$ & $\underline{20.9}_{0.1}$ & $29.3_{0.8}$ & $\underline{20.8}_{0.4}$ & $12.5_{0.5}$ & $43.3_{0.5}$ \\
Few-shot FT & $38.4_{1.8}$ & $78.6_{14.7}$ & $38.4_{3.8}$ & $29.9_{4.7}$ & $62.6_{3.6}$ & $19.4_{1.9}$ & $60.5_{13.6}$ & $17.3_{1.8}$ & $10.9_{1.2}$ & $57.7_{6.5}$ \\
CaLF & $\underline{40.3}_{0.2}$ & $\underline{84.0}_{3.3}$ & $\underline{41.7}_{1.2}$ & $\underline{34.5}_{0.5}$ & $\underline{81.5}_{2.5}$ & $20.4_{1.5}$ & $\underline{62.7}_{4.6}$ & $18.4_{2.1}$ & $\underline{13.1}_{0.7}$ & $\underline{73.1}_{4.2}$ \\
Ours & $\textbf{41.0}_{0.3}$ & $\textbf{84.1}_{1.1}$ & $\textbf{43.1}_{1.5}$ & $\textbf{36.5}_{0.3}$ & $\textbf{84.9}_{0.8}$ & $\textbf{22.0}_{0.6}$ & $\textbf{63.5}_{2.9}$ & $\textbf{21.4}_{1.1}$ & $\textbf{15.7}_{0.6}$ & $\textbf{76.3}_{2.1}$ \\
\bottomrule
\end{tabular}
\caption{Main results on ALCE benchmark (ASQA and ELI5). We report the mean and standard deviation measured across 3 different random seeds. For each dataset, we use Citation F1 to measure citation quality, EM Recall to assess overall correctness, Correct. in P to evaluate the proportion of correct information grounded in the retrieved passages, Rouge-L to assess textual similarity, and MAUVE to evaluate fluency. Bold indicates the best performance, while underline denotes the second-best.}
\label{main_results}
\end{table*}

\section{Experiments}
\subsection{Datasets \& Metrics}
We conduct experiments on the ALCE citation benchmark \cite{gao2023enabling} for long-form question answering, focusing primarily on the ASQA and ELI5 datasets. In ALCE, the number of reference documents provided for training is $\mathcal{D} = 4$. We adopt the evaluation protocol from \cite{gao2023enabling}, which includes correctness (measured by EM Recall), fluency (measured by MAUVE), and citation F1 (assessed using an NLI-trained T5-11B model). In addition, we report ROUGE-L scores and the passage-grounded correctness (Correct. in P) metric introduced in \cite{aly2024learning}, which evaluates whether the response is supported by the retrieved documents P, disregarding factual content potentially memorized by the language model.

\subsection{Experimental Setup}
To better reflects realistic deployment scenarios, we follow few-shot learning recommendations \cite{alex2021raft} and omit a separate validation set for hyperparameter tuning. Due to computational constraints, we adopt LoRA \cite{hu2022lora} for parameter-efficient fine-tuning. We use AlignScore as our FCM. The threshold $\tau_{cite}$ and $\tau_{ans}$ is set to 0.8 and 0.45. Notably, AlignScore differs from the citation evaluation model used in our final evaluation in both architecture and training data. In terms of knowledge sources,  ASQA uses Wikipedia as its knowledge source, while ELI5 relies on CommonCrawl. For fair comparison, we use the same retriever as employed by baseline models. In the behavioral pattern learning stage, we perform 1 epoch of warm-up followed by 3 epochs of online self-reflective bootstrapping with a learning rate of 1e-4. In the RL phase, we set $\tau_{\text{cite}}^{\text{attempt}}$ and $\tau_{\text{ans}}^{\text{attempt}}$ to 0.7 and 0.4. The learning rate is set to 5e-6, batch size to 16, KL coefficient to 0.1, and train for 3 epochs.

\begin{table*}[ht]
\setlength{\tabcolsep}{1mm}
\centering
\resizebox{\textwidth}{!}{
\begin{tabular}{l|ccccc|l|ccccc}
\toprule
\textbf{Method} & Similarity & Fluency & Correct & Correct & Citation & \textbf{Method} & Similarity & Fluency & Correct & Correct & Citation \\
($Source\rightarrow Target$) & Rouge-L & MAUVE & EM Rec. & . in P & F1 & ($Source\rightarrow Target$) & Rouge-L & MAUVE & EM Rec. & . in P & F1 \\
\midrule
Self-RAG 7B & 35.7 & 74.3 & 30.0 & -   & 67.3  & Self-RAG 7B & 16.9 & 32.6 & 9.7  & 5.4  & 27.6 \\
Zero-Shot\textit{($\rightarrow A$)} & 39.0 & 78.9 & 39.5 & 31.6 & 5.7 & Zero-Shot\textit{($\rightarrow E$)} & \textbf{21.3} & 35.0 & 22.2 & 12.6 & 10.4 \\
Few-shot FT\textit{($E\rightarrow A$)} & 39.7 & \textbf{90.1} & 38.5 & 31.4 & 71.7 & Few-shot FT \textit{($A\rightarrow E$)} & 20.9 & \textbf{41.1} & 19.7 & 10.6 & 40.4 \\
Calf{($E\rightarrow A$)} & 40.1 & 86.6 & 40.0 & 33.2 & 79.5 & Calf\textit{($A\rightarrow E$)} & 21.2 & 31.3 & 20.4 & 12.5 & 57.3 \\
Ours\textit{($E\rightarrow A$)} & \textbf{40.5} & 87.2 & \textbf{42.3} & \textbf{34.7} & \textbf{81.3} & Ours\textit{($A\rightarrow E$)} & 21.0 & 32.4 & \textbf{23.6} & \textbf{13.4} & \textbf{59.1} \\
\bottomrule
\end{tabular}
}
\caption{Zero-shot domain transfer evaluation results on ASQA (A) and ELI5 (E). We report the results using Mistral-Orca-7B.}
\label{transfer}
\end{table*}

\subsection{Baselines}
We compare our method against a range of strong baselines. First, we include in-context prompting and few-shot fine-tuning approaches based on the same instruction-tuned LLMs, trained on the few-shot citation dataset $D$. We further compare with the state-of-the-art model CaLF \cite{aly2024learning}, which iteratively trains on filtered, self-generated data. Additionally, we evaluate against powerful in-context prompting baselines from \cite{gao2023enabling}, including ChatGPT (gpt-3.5-turbo-0301) and GPT-4 (gpt-4-0613), which use large-scale parameters.In-context prompting uses two randomly sampled demonstrations. We also consider several recent citation-aware models: AGREE \cite{ye2023effective}, based on PaLM 2; Self-RAG 7B \cite{asai2024self}, built on the open-source Llama2 backbone; and Blueprint (BP) \cite{fierro2024learning}, which is based on T5-3B.

\subsection{Main Results}
Table~\ref{main_results} presents the results of the in-domain experiments. Overall, FineRef consistently achieves the highest or highly competitive performance across all evaluation metrics compared to baseline methods. Our approach significantly outperforms larger-scale models such as ChatGPT and GPT-4 on all metrics. Across different base LLMs (LLaMA2-7B-chat and Mistral-Orca-7B), FineRef substantially surpasses both in-context and few-shot fine-tuning approaches, with up to +22.3 improvement in Citation F1 over the few-shot FT baseline on ASQA. Compared to the state-of-the-art model, CaLF, FineRef demonstrates notable improvements in both citation and correction performance. While CaLF achieves better citation quality than GPT-4, its improvement in correction is marginal (e.g., only +0.4 EM on ASQA). In contrast, FineRef yields consistent gains over GPT-4 in both citation and QA quality, with improvements of +2.8 EM on ASQA and +7.2 EM on ELI5. These results highlight the effectiveness of our fine-grained reflection mechanism in jointly improving citation accuracy and answer quality by explicitly modeling different types of citation errors.

\subsection{Domain Transfer}
Table~\ref{transfer} presents the results of our zero-shot transfer experiments, where models trained on one domain are directly evaluated on a different target domain without further fine-tuning.across all source-to-target settings, our method (based on MistralOrca-7B) consistently outperforms Self-RAG, Zero-Shot, Few-shot Fine-tuning, and the state-of-the-art method Calf in both citation quality and correction performance. On the ASQA target domain, our approach (trained on ELI5) achieves notable improvements in EM Recall (+2.3), Correct in Passage (+1.5), and Citation F1 (+1.8) compared to the previous state-of-the-art, Calf. Similarly, on the ELI5 target domain, our model (trained on ASQA) maintains superior performance in EM Recall (23.6) and Citation F1 (59.1), outperforming Calf by margins of +3.2 and +1.8 respectively. This demonstrates that our approach, which explicitly trains the model to reflect on different types of citation errors, leads to better generalization across domains in both citation and answer quality.

\begin{table}[ht]
\centering
\resizebox{\columnwidth}{!}{\begin{tabular}{l|ccc}
\toprule
\multirow{2}{*}{\textbf{Method}} & Correct & Correct & Citation \\
& EM Rec. & . in P & F1 \\
\midrule
Ours & 43.4 & 36.3 & 85.2 \\
\midrule
wo reflection & $38.2_{(-5.2)}$ & $32.6_{(-3.7)}$ & $79.4_{(-5.8)}$ \\
wo IR error type & $39.9_{(-3.5)}$ & $33.3_{(-3.0)}$ & $82.1_{(-3.1)}$ \\
coarse-grained refl. & $40.2_{(-3.2)}$ & $33.8_{(-2.5)}$ & $81.1_{(-3.1)}$ \\
\midrule
wo RL & $40.8_{(-2.6)}$ & $33.9_{(-2.4)}$ & $80.6_{(-4.6)}$ \\
wo gain reward & $42.0_{(-1.4)}$ & $32.9_{(-3.4)}$ & $81.6_{(-3.6)}$ \\
wo online boost. & $41.1_{(-2.3)}$ & $34.2_{(-2.1)}$ & $81.5_{(-3.7)}$ \\
\bottomrule
\end{tabular}}
\caption{Results for ablation study. We report the results on ASQA dataset using Mistral-Orca-7B.}
\label{ablation}
\end{table}
\subsection{Analysis}
\paragraph{Ablation.} To better understand the underlying mechanisms of our method, we conduct a series of ablation studies to examine the contribution of each component. Table~\ref{ablation} presents the results on the ASQA dataset using the Mistral-Orca-7B model. We first analyze the role of the \textbf{reflection} module. Removing the reflection behavior results in a substantial drop in both EM Recall (-5.2) and Citation F1 (-5.8), indicating that the reflection step plays a vital role in improving both citation accuracy and answer quality.

We further investigate the impact of excluding the identification of \textbf{irrelevant} citations during reflection. This leads to notable declines in EM Recall (-3.5), highlighting the importance of modeling irrelevant citation errors to maintain question-answering capability. Additionally, we compare our fine-grained reflection strategy to a \textbf{coarse-grained alternative} that uses GPT-4 to assess the overall correctness of the attempt. This approach yields inferior performance across all quality metrics, confirming the necessity of our fine-grained reflection formulation.

We also evaluate the effect of different components in the training strategy. Removing the \textbf{RL} stage leads to notable performance drops (-2.6 EM and -3.6 Citation F1), highlighting the importance of self-exploration in strengthening reflective and corrective behaviors. The \textbf{gain reward} also contributes positively by preventing behavioral collapse and encouraging higher-quality correction generations. Finally, ablating the \textbf{online self-reflective bootstrapping (online boost.)} leads to declines in all metrics, as it weakens the model’s initial behavior pattern learned during SFT, thus limiting the improvement achievable during the RL phase.

\begin{table}[ht]
\centering
\begin{tabular}{l|ccc}
\toprule
\multirow{2}{*}{\textbf{Method}} & Correct & Correct & Citation \\
& EM Rec. & . in P & F1 \\
\midrule
First Attmpt & 38.6 & 30.3 & 77.4 \\
Correction & 43.1 & 36.5 & 84.7 \\
$\Delta$ & +4.5 & +6.2 & +7.3 \\
\bottomrule
\end{tabular}
\caption{Performance comparison of first attempt and correction. We report the results on ASQA using Mistral-Orca-7B.}
\label{improvement}
\end{table}

\begin{figure}[htbp]
\centering
    \subfigure[FCM+Reranker Reflection]{
       \centering
        \includegraphics[width=0.45\columnwidth]{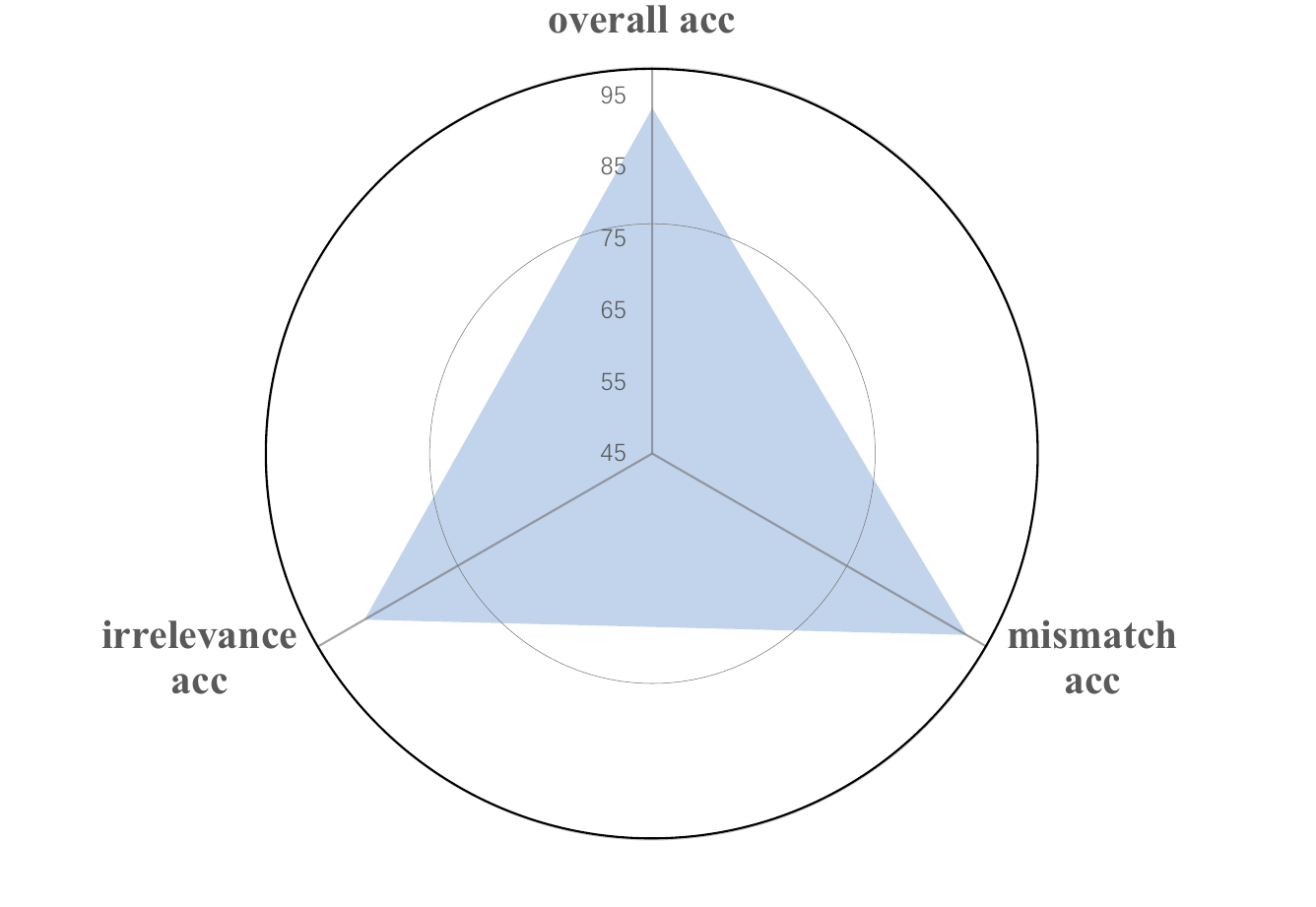}
    }\subfigure[Self-Reflection]{
        \centering
        \includegraphics[width=0.45\columnwidth]{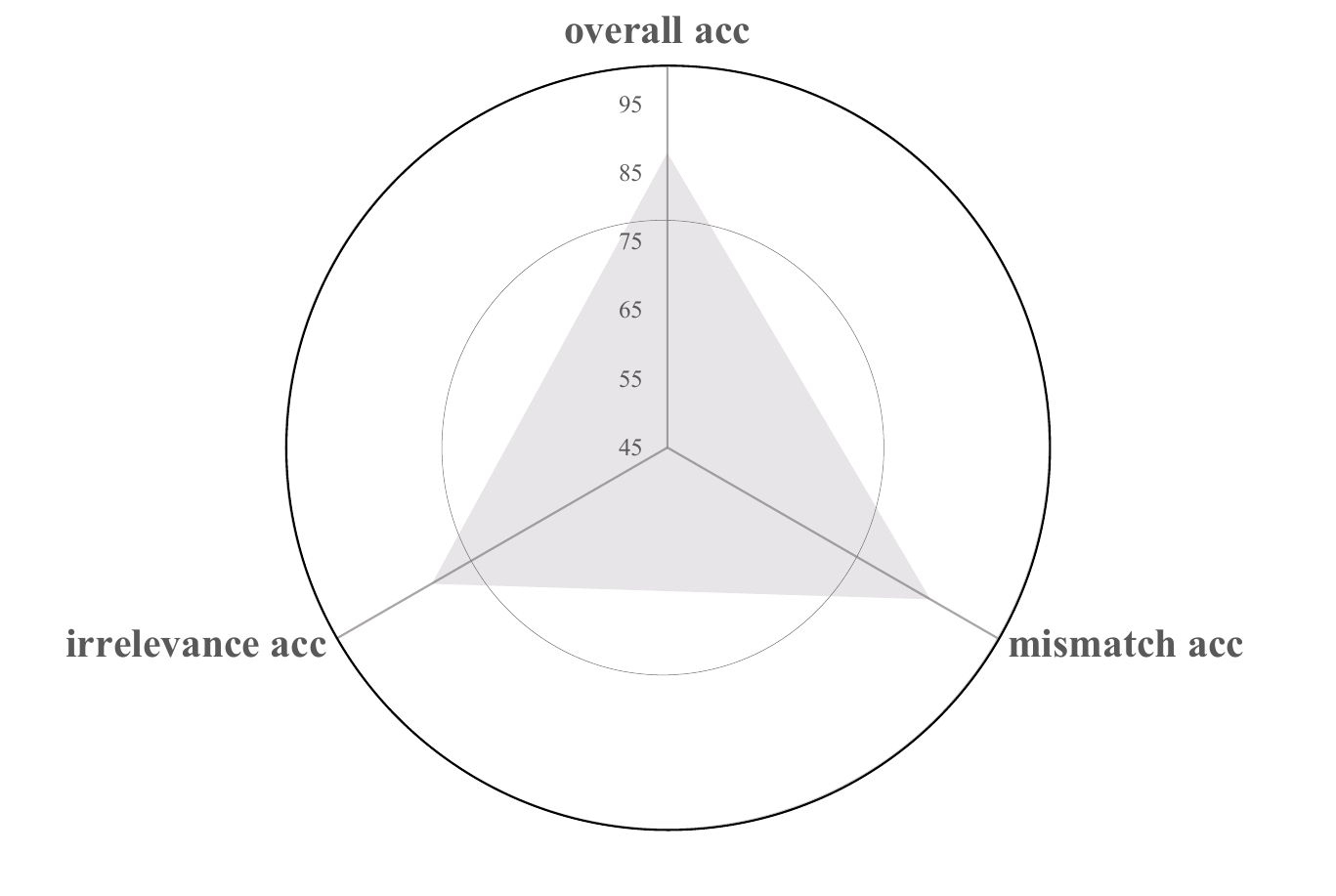}
    }\caption{Accuracy of self-reflection\label{reflect_acc}}
\end{figure}

\paragraph{Correction Improvement.} To assess the effectiveness of the correction behavior in enhancing initial generation quality, Table~\ref{improvement} presents the performance improvements of the model after applying correction relative to the first attempt. Notably, all evaluation metrics show consistent gains (Citation F1 +7.3, EM Recall +4.5 and Correct in P +6.2). These results highlight that the model's initial attempt often falls short in citation and answer quality, whereas the subsequent correction—guided by fine-grained self-reflection—substantially enhances the reliability and accuracy of the generated response.

\paragraph{Reflection Accuracy Analysis.} To evaluate the accuracy of reflection, we conduct a two-step analysis. First, we need to ensure that the FCM and reranker produce reliable reflection signals. We sample a subset of 400 test samples from ASQA to evaluate the accuracy of reflection labels constructed by FCM and reranker. Two human evaluators are tasked with verifying whether the predicted error types (e.g., mismatch or irrelevance) are accurate. As shown in Figure~\ref{reflect_acc}(a), the reflection signals based on FCM and reranker exhibit high accuracy (over 90\%) for reflecting on different types of errors. Next, we use the FCM and reranker as gold reflection signals, and evaluate the accuracy of the model's self-generated reflections. The evaluation across all test samples of ASQA is as illustrated in Figure~\ref{reflect_acc}(b). This measures the model's ability to correctly identify citation errors and improve its own performance through self-reflection.

\begin{figure}[htbp]
\centering
    \subfigure[Citation F1]{
       \centering
        \includegraphics[width=0.45\columnwidth]{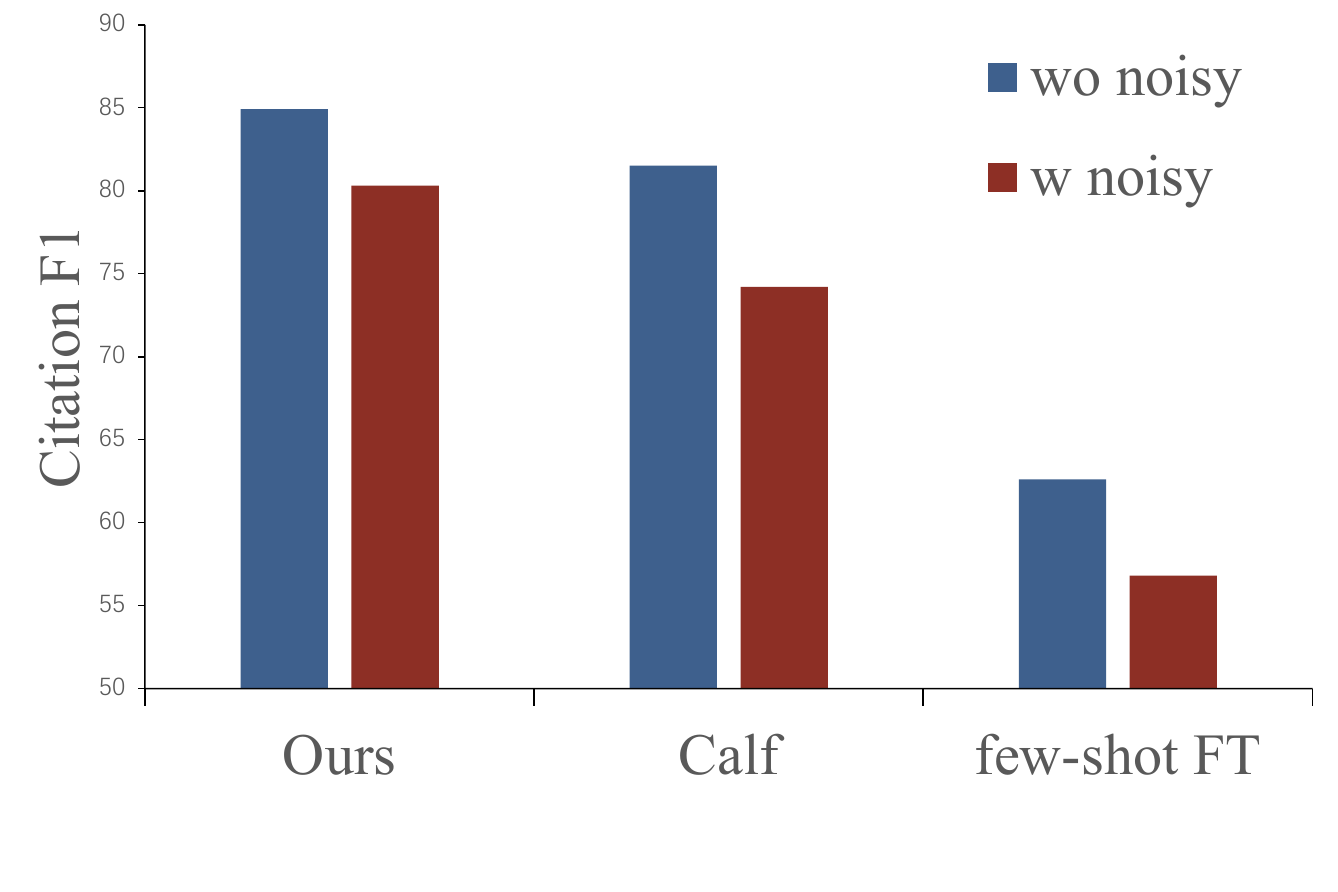}
    }\subfigure[EM Recall]{
        \centering
        \includegraphics[width=0.45\columnwidth]{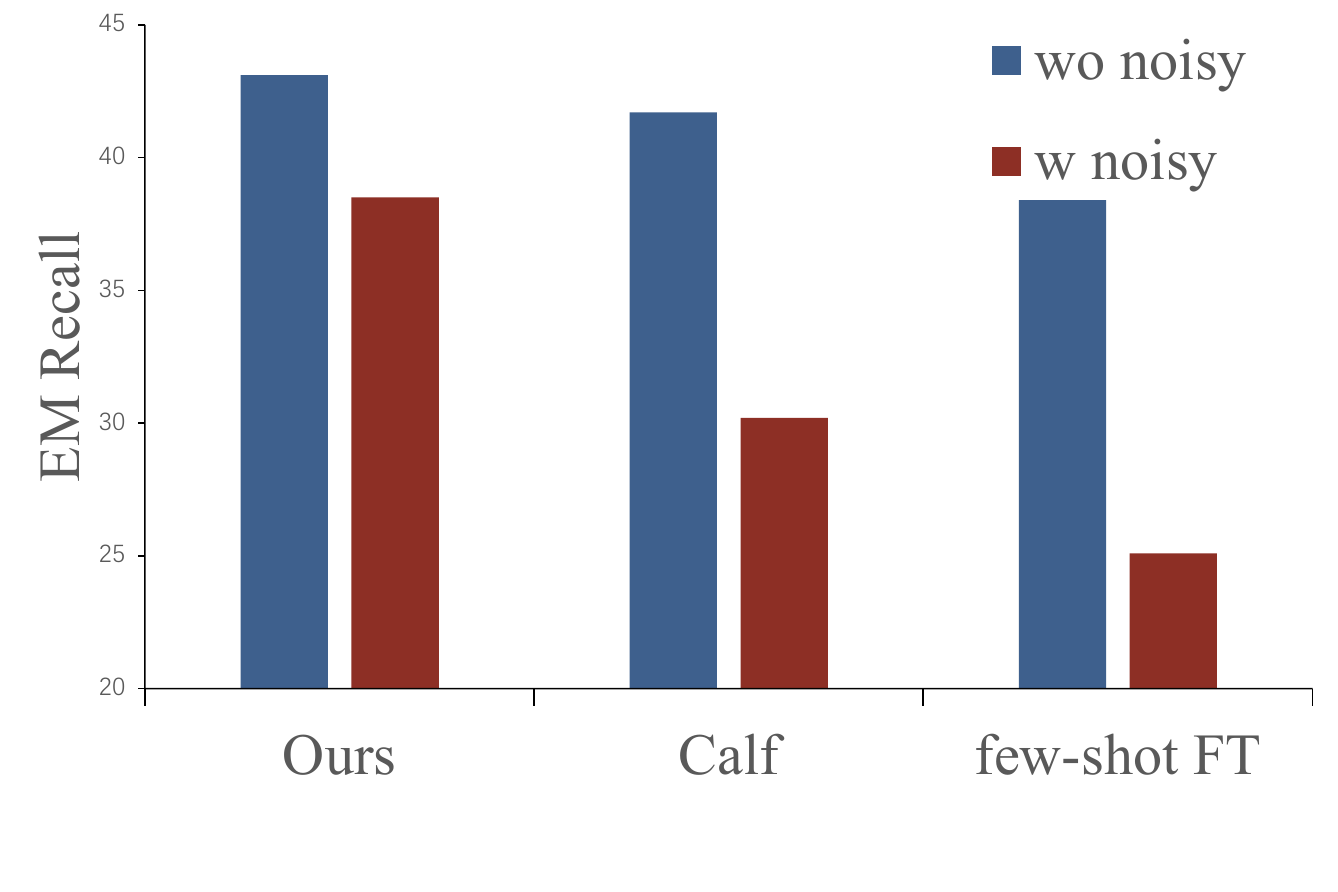}
    }\caption{Performance in the scenario with noisy passages\label{noisy}}
\end{figure}

\vspace{-1.0em}

\paragraph{Robustness under Noisy Retrieval.} To evaluate the robustness of our method under realistic retrieval noise conditions, we introduce semantically fluent yet contextually irrelevant distractor passages, simulating common failure modes of open-domain retrievers. Concretely, for each input question, we randomly replace one of the retrieved passages with such a distractor. This setup emulates real-world retrieval systems that often return noisy or irrelevant content. 

We compare FineRef against the state-of-the-art CaLF method and a Few-shot fine-tuning baseline. As shown in Figure~\ref{noisy}, the performance of both CaLF and Few-shot fine-tuning degrades substantially in the presence of distractor passages, particularly EM Recall. In contrast, FineRef consistently maintains a significant lead in both Citation F1 and EM Recall. These results validate the effectiveness of our approach in enhancing model robustness through explicit citation error modeling and fine-grained reflection. 

\section{Related Work}
\subsection{Text Generation with Citations}
Generating text with citations enhances LLM verifiability and mitigates hallucinations \cite{gao2023enabling, li2023helma}. Early practices integrated LLMs with commercial search engines\footnote{https://www.perplexity.ai/}\footnote{https://www.bing.com/new} , complicating citation evaluation and prompting the development of benchmarks like ALCE \cite{gao2023enabling}. Most work adopts a “single-pass generation” paradigm, allowing LLMs to directly produce cited answers from retrieved documents. To improve this process, methods such as in-context learning with high-quality exemplars \cite{gao2023enabling,liu2023evaluating}, post-hoc citation editing \cite{bohnet2022attributed}, and Reinforcement Learning from Human Feedback (RLHF) \cite{menick2022teaching, thoppilan2022lamda, nakano2021webgpt} have been proposed. Weakly-supervised strategies like CaLF \cite{aly2024learning} further address data scarcity by using Factual Consistency Models (FCMs) to filter diverse candidate answers for training. Despite this progress, key limitations persist: current methods prioritize citation \textit{fidelity} over \textit{relevance} compromises QA performance and robustness under noisy real-world conditions, and hinders the model's ability to generate optimal responses in long-form tasks requiring multiple citations. FineRef introduces a fine-grained error reflection mechanism that explicitly trains the model to identify and correct both “mismatch” and “irrelevance” citation errors, enhancing its overall performance and robustness.

\subsection{LLM Reasoning and Self-Reflection}
Self-reflection is a critical mechanism for improving complex reasoning abilities and reliability of LLMs \cite{madaan2023self,paul2023refiner}, enabling models to evaluate and iteratively refine their responses in a human-like manner. Recent studies have explored various approaches to enhance these capabilities during post-training \cite{saunders2022self, rosset2024direct, kumar2024training}. Enabling LLMs to perform effective self-verification and self-correction is a promising solution for achieving robust reasoning, as direct prompting for such behaviors is often suboptimal \cite{huang2023large, yixing2024chain, zhang2024understanding}. RL has also proven effective in enhancing LLM reasoning \cite{setlur2024rl, ouyang2022training}, with research focusing on actor-critic frameworks \cite{havrilla2024teaching,tajwar2024preference} and the design of accurate reward models to guide the learning process \cite{lightman2023let}. Some work \cite{ma2025s}, has combined these ideas, using RL to teach models how to self-verify and self-correct. Existing self-reflection mechanisms are often too coarse-grained and rely on unreliable self-generated feedback, limiting their effectiveness. FineRef introduces a fine-grained, controllable approach that uses external specialized and lightweight models to precisely reflect errors at the citation level.

\section{Conclusion}
In this paper, we propose FineRef, a noval training framework that enables LLMs to perform fine-grained self-reflection for identifying and correcting both mismatch and irrelevance citation errors. FineRef explicitly models the full attempt–reflect–correct process through supervised learning with fine-grained and controllable reflection signals and enhances this capability via process-level RL with multi-dimensional rewards design. Extensive experiments demonstrate that FineRef not only substantially improves citation fidelity and answer correctness but also exhibits strong robustness across domains and under retrieval noise. These results highlight the effectiveness of incorporating fine-grained self-reflection into training, representing a promising step toward building more reliable, interpretable, and trustworthy language models.

\section{Acknowledgments}
This research is supported by Artificial Intelligence-National Science and Technology Major Project 2023ZD0121200.

\bibliography{aaai2026}

\end{document}